\documentclass[preprint,aps,eqsecnum,showkeys,showpacs]{revtex4}
\usepackage{amsmath}

\begin{document}

\title{Position and Spin Operators, Wigner Rotation and the Origin of Hidden Momentum Forces}
\author{R. F. O'Connell\thanks{
oconnell@phys.lsu.edu}}
\affiliation{Department of Physics and Astronomy, Louisiana State University, Baton
Rouge, LA 70803-4001 USA}
\date{\today }

\begin{abstract}
Using a position operator obtained for spin $\frac{1}{2}$ particles by the present author and Wigner, we obtain a quantum relativistic result for the hidden momentum force experienced by particles with structure.  In particular, our result applies to the hidden magnetic forces manifest in some problems of electromagnetism. We also discuss spin and orbital angular momentum operators, as well as Wigner rotation.
\end{abstract}

\maketitle

\section{Introduction}

A striking feature of quantum mechanics is the fact that the momentum $\vec{P}$ and the velocity $\vec{v}$ are not necessarily proportional to each other. As noted by the present author and Wigner \cite{oconnell77}, "-- although $\vec{P}$ has a natural definition - it is the generator of spatial translations of the state vector \cite{wigner39, bargmann48} -- this is not so for the operator $\vec{v}$." This motivated us to examine more generally the relation between momentum and velocity, the goal being to show under what circumstances one obtains the operator equivalent of the relation. 

\begin{equation}
\vec{v}=\vec{P}/P_{0}, \label{hmf1}
\end{equation}
where (in units $c=\hbar$ =1)

\begin{equation}
P_{0}=(m^{2}+\vec{P}^{2})^{1/2} \\
\equiv (m^{2}+P^{2})^{1/2}, \label{hmf2}
\end{equation}
for the relation between the momentum $\vec{P}$ and velocity $\vec{v}$ in \underline{relativistic classical} mechanics. In addition, if the velocity operator $\vec{v}$ is defined as

\begin{equation}
\bar{v}=\frac{d\vec{q}}{dt} \label{hmf3}
\end{equation}
such that its expectation value is given by

\begin{equation}
\left\langle\vec{v}\right\rangle =\frac{d \left\langle\vec{q}\right\rangle}{dt}, \label{hmf4}
\end{equation}
where $\vec{q}$ is the position operator, then it does depend on the definition of the position, that is on the form chosen for the position operator.  We then showed that the position operator was that of Newton and Wigner \cite{newton49} for \underline{spin $0$} particles, leading to the result \cite{oconnell77}

\begin{equation}
\left\langle \vec{q}(t)\right\rangle =\left\langle \vec{q}(0)\right\rangle + t\left\langle \vec{v}\right\rangle. \label{hmf5}
\end{equation}
However, equation (\ref{hmf5}) will not be valid for a particle with spin.

In fact, even in relativistic classical mechanics, when spin is included, the relation between $\vec{P}$ and $\vec{v}$ is no longer necessarily simple nor unique. Moller \cite{moller72} pointed out that in special relativity, a particle with structure and "spin" (its angular momentum vector in the rest system) is subject to a spin supplementary condition, which "-- expresses in a covariant way that the proper center of mass is the center of mass in its own rest system $(K^{0})$ --" and that "-- the difference between simultaneous positions of the centre of mass in $K$ (obtained from $K^{0}$ by a Lorentz transformation with velocity $(\vec{v})$ and the proper centre of mass (in $K^{0}$) --" is

\begin{equation}
\Delta\vec{r}=\frac{\vec{S}\times\vec{v}}{mc^{2}} \label{hmf6}
\end{equation}
where $(\vec{S})$  is the spin and $m$ is the rest mass.  Equation (1.6) arose in Corben's analysis of the motion of a free gyroscope in the absence of external forces or torques \cite{corben68} and this special relativistic effect also played a role in the calculation of spin precession in a general relativistic field by Barker and the present author \cite{barker70, barker74}. In essence, it is related to the fact that, in special relativity, there are two rest systems for the particle, zero velocity and zero momentum, reflecting the choice of spin supplementary conditions and the fact that only in special cases are the velocity and momentum proportional to each other.  We refer to an extensive review for more details \cite{oconnell10}.  If one neglects the second-order terms in $\vec{S}$, which arise because of the (unspecified) relation between the velocity $\vec{v}$ and momentum $\vec{P}$ and if $\vec{v}$ $\ll$ $c$, then one can simply take the time derivative of (\ref{hmf6}) to obtain the so-called hidden momentum.

\begin{equation}
\Delta\vec{P}=\frac{\vec{S}\times\vec{F}}{mc^{2}} \label{hmf7}
\end{equation}
where $\vec{F}$ is the external force \cite{oconnellpress,gralla12}. This is an \underline{approximate} result.

We note that Moller's result is \underline{classical} and depends on the choice of the spin supplementary condition and consequently the same remark applies to equation (\ref{hmf7}). Hence, we are motivated to provide a quantum mechanical derivation.  

Thus, we considered the case of a spin $\frac{1}{2}$ particle and we found that a new position operator, $Q$ say, was required \cite{oconnell78}, where (after reinserting the $\hbar$ which was taken to be unity in \cite{oconnell78}) we obtained

\begin{equation}
\vec{Q}=\vec{q}+\hbar\left(\vec{P}\times\vec{\sigma}\right) /P^{2}, \label{hmf8}
\end{equation}
where $\vec{\sigma}$ is the Pauli spin operator. 

This is a key quantum mechanical operator result, which was obtained rigorously in \cite{oconnell78}. It is unique. 

Next, just as we successfully replaced $\frac{\hbar}{2}\vec{\sigma}$ by the classical spin $\vec{S}$ in our discussion of classical spin precession \cite{barker70}, we now write

\begin{eqnarray}
\vec{Q }&=& \vec{q}+2\left(\frac{\vec{P}\times\vec{S}}{P^{2}}\right) = \vec{q}+2\left(\frac{\vec{v}\times\vec{S}}{P_{0}{v^2}}\right) \nonumber\\
&\equiv& \vec{q}+\Delta\vec{q}, \label{hmf9}
\end{eqnarray}
as applying to any spin angular momentum.

Thus, our result $\Delta\vec{q}$ is our generalization of the result, given in (\ref{hmf5}). We should emphasize that our derivation is both quantum mechanical and relativistic.  As in the case of the original Newton-Wigner derivation, our analysis pertains to an arbitrary positive energy state of the system. Thus, it is more complementary to the second-order equation for the electron \cite{feynman58} than to the Dirac equation.

"Hidden velocity" which clearly depends on the acceleration of the particle demonstrates that we are continually moving to a different Lorentz frame.  In fact, this feature is analogous to Thomas precession or Wigner rotation \cite{wigner39} where "-- the electron's rest frame of coordinates is defined as a co-moving sequence of inertial frames whose successive origins move at each instant with the velocity of the electron" \cite{jackson98}.

In order to compare the operators $\Delta$$\vec{q}$ with the classical quantity $\Delta$$\vec{v}$ it is necessary to take the expectation value of $\Delta$$\vec{q}$. However, this will necessitate consideration of the particular system being analyzed, a point recently emphasized by Bauke et al., \cite{oconnell77, bauke14} in their effort to distinguish experimentally between a variety of relativistic spin operators in various electromagnetic environments. In this context, it is also of interest to note that, since total angular momentum is a constant of the motion, the new position operator (\ref{hmf8}) we have introduced also implies a change in the corresponding spin operator and, concomitantly, a change in the orbital angular momentum operator, such that the original $l$= 0 now became $l$= 1 \cite{oconnell78}. This could also explain why Bohr's result for the ground state of hydrogen was $l$=0 instead of the expected $l$=1.

Also, apart from the unimportant sign, the appearance of $\frac{2}{P_{0}{v^2}}$ in $\Delta$$\vec{q}$ compared to $\frac{1}{mc^{2}}$ in (\ref{hmf7}), is surprising and warrants future investigation. On the other hand, the important factor $(\vec{v}\times\vec{S})$ is common to both cases.

\section{DISCUSSION AND CONCLUSIONS}

All of the above results depend only on special relativity. However, an important application is to the particular case of electrodynamics, since the expression for a magnetic moment $\vec{M}$ is derived from either the spin of a particle or from a steady current (bodies with structure in both cases).  Thus, in this case, with $\Delta\vec{P}\approx$ $m$($\Delta\vec{v})$, consistent with the neglect of $(spin)^2$ terms as in the derivation of (\ref{hmf8}),

\begin{equation}
\Delta\vec{P} =\frac{\vec{S}\times\vec{a}}{c^{2}}, \label{hmf13}
\end{equation}
where  $\vec{a}$ is the acceleration.  If the magnetic moment is interacting with a pointlike electric charge $e$, then the electric field $\vec{E}$ created gives rise to a force $e\vec{E}$ so that

\begin{equation}
\Delta\vec{P}=k_{1}\frac{\vec{M}\times\vec{E}}{c^{2}}, \label{hmf14}
\end{equation}
where $k_{1}$ is a constant.  This is the quantum relativistic generalization of the familiar non-relativistic result, given by $\left(\vec{E}\times\vec{M}\right)/c^{2}$ \cite{griffiths99,griffiths12} and many authors \cite{vanzella13} to explain the results of Mansuripur \cite{mansuripur12}. However, it appears that the essence of the dispute is connected with the choice of coordinate systems. While it is true to say that the laws of physics, including the Lorentz force law, are the same in all inertial systems, the point here is that, in the presence of an external force, the inertial frames are continually changing.

This work was partially supported by the National Science Foundation under grant no. ECCS-1125675.

\end{document}